# X-VECTOR BASED VOICE ACTIVITY DETECTION FOR MULTI-GENRE BROADCAST SPEECH-TO-TEXT

*Misa Ogura, Matt Haynes*

BBC Research & Development, UK

## ABSTRACT

Voice Activity Detection (VAD) is a fundamental pre-processing step in automatic speech recognition. This is especially true within the broadcast industry where a wide variety of audio materials and recording conditions are encountered. Based on previous studies which indicate that x-vector embeddings can be applied to a diverse set of audio classification tasks, we investigate the suitability of x-vectors in discriminating speech from noise. We find that the proposed x-vector based VAD system achieves the best reported score in detecting clean speech on AVA-Speech, whilst retaining robust VAD performance in the presence of noise and music. Furthermore, we integrate the x-vector based VAD system into an existing STT pipeline and compare its performance on multiple broadcast datasets against a baseline system with WebRTC VAD. Crucially, our proposed x-vector based VAD improves the accuracy of STT transcription on real-world broadcast audio.

## 1. INTRODUCTION

Large broadcasters such as the BBC produce thousands of hours of AV content each day. Accurate automatic Speech-to-Text (STT) systems can unlock more value from such content, aiding discoverability, reusability and accessibility. In previous decades, the costs, lack of specialisms and low accuracy of generating STT transcripts on broadcast audio have impeded large scale adoption. However, advancement in accuracy and efficiency, as well as development of open-source toolkits such as Kaldi ASR [1], have enabled the use of STT as a practical tool for broadcasters and journalists [2]. A series of academic challenges based on broadcast datasets in multiple languages [3][4][5] demonstrates continuing interests from both the academia and the industry in making STT systems work better on real-world broadcast audio. One of the many challenges in achieving a reliable broadcast STT system is voice activity detection (VAD).

VAD refers to the task of distinguishing active speech from non-speech. Detecting active speech signals in broadcast content poses particular difficulties due to the presence of non-speech vocal sound, extended silences, environmental noise and background music that often coincide with speech. As it serves as an essential pre-processing step, developing a robust VAD system is of particular importance to numerous downstream applications such as STT, speaker diarisation and speaker recognition. Earlier VAD approaches include energy-based thresholding [6], spectral analysis [7][8] and pitch detection [9]. Following the work of Sohn et.al. [10], where parameter estimation was applied using the Gaussian distribution, several statistical methods have been proposed [11][12][13]. More recently, machine learning-based systems have achieved state-of-the-art performance by considering VAD as a frame-level speech/noise classification problem. Methods utilising support vector machines (SVM) [14][15][16] and deep neural networks (DNN) [17][18][19][20] are amongst the most actively studied.

X-vectors [21] are 512-dimensional audio embeddings extracted from a DNN [22] trained to classify speakers in the training data. The x-vector system was shown to significantly outperform i-vector baseline systems in speaker recognition benchmarks [21]. Application of x-vectors ranges from speaker diarisation [23][24], speaker recognition [25][26], language detection [27] to acoustic scene classification [28], emotion classification [29] and early detection of Parkinson's Disease [30][31].

The success of applying x-vectors to a diverse set of audio classification tasks led us to reason that x-vectors can be used as features to train a VAD system. To this end, we propose that 1) X-vectors contain latent information to differentiate speech from noise in audio signals and hence 2) it is possible to use x-vectors as features to train a VAD system which, when applied upstream, 3) improves the performance of an STT system. In this paper, we present our work to address these propositions, with a particular focus on developing a robust and accurate VAD system that can detect speech reliably in the presence of a variety of noise, and its application to automatic STT transcription of broadcast audio.

## 2. DATASETS AND DATA PREPARATION

In this section, we introduce datasets used for training and evaluation, as well as a protocol for x-vector extraction.

## 2.1. Multi-Genre Broadcast (MGB)

MGB is a broadcast dataset to train acoustic and language models [3]. The training set (`train.full`) is accompanied by lightly supervised transcripts based on the original BBC subtitles, whereas the STT evaluation set (`eval.std`) is provided along with manually annotated transcripts. We utilised `train.full` as a source of speech data for our study, as it provides a collection of whole TV shows drawn from diverse genres, with variety of recording conditions and qualities. Original utterances were decoded and aligned using an in-house STT system to obtain more granular word-level time codes. New ground truth was generated from these time codes to increase the quality of segment boundaries and to remove any longer gaps of silence, resulting in the increased number of aligned segments albeit a shorter total duration of aligned speech (`realigned.train.full`). A summary of data subsets is shown in Table 1.

## 2.2. Audio Set

Audio Set is a large-scale dataset consisting of over 2 million manually annotated 10 second segments from YouTube clips [32]. Segments are labelled with one or more audio events that are structured into an ontology of 632 classes. As the Audio Set ontology provides comprehensive coverage of real-world sounds from all domains, it makes an ideal source of everyday noise. From the balanced training subset, we created a noise dataset (Audio Set Noise) by excluding segments labelled with the "Speech" class (ontology identifier: /m/09x0r). Audio Set Noise contains 14,516 segments.

## 2.3. BBC-VAD

We curated a new, densely labelled VAD dataset by combining x-vector embeddings extracted from Audio Set Noise (except for 1,052 segments with erroneous labels which were manually identified and removed) and comparable amount of speech segments from MGB `realigned.train.full`. BBC-VAD consists of 13,464 labelled noise segments (36.9 hours) and 16,464 labelled speech segments (39.8 hours). Table 2 shows a breakdown of training (`bbc-vad-train`) and testing (`bbc-vad-eval`) subsets in terms of audio duration. The noise portion represents 530 distinct Audio Set Ontology events, and the speech portion represents speech data sampled across 2,164 BBC programmes, making BBC-VAD a highly diverse dataset with everyday noise and broadcast speech.

## 2.4. AVA-Speech

AVA-Speech is a speech activity dataset which consists of audio segments taken from minutes 15-30 from 160 YouTube clips [33]. In comparison to other VAD datasets where utterances are taken from limited domains such as telephone conversations [34] or meeting recordings [35], AVA-Speech contains a diverse set of speakers engaging in natural conversation and various acoustic conditions, with roughly equal amounts of speech and noise data. Over 70% of its speech data is accompanied by background noise or music, providing an excellent approximation of noise in real-word broadcast media. At the time of our experiments, 27% of the dataset was not available for analysis.

| Subset | # shows | Total dur (h) | Aligned speech (h) | # aligned segments |
|---|---|---|---|---|
| train.full | 2193 | 1580 | 1197 | 635,827 |
| eval.std | 16 | 11 | N/A | N/A |
| realigned.train.full | 2192 | 1579 | 808 | 906,590 |

**Table 1:** Aggregate statistics over training (`train.full`), and evaluation (`eval.std`) subsets from the MGB dataset, as well as realigned training dataset (`realigned.train.full`).

| Subset | Speech dur (h) | Noise dur (h) | Total dur (h) |
|---|---|---|---|
| bbc-vad-train | 35.9 | 33.2 | 69.1 |
| bbc-vad-eval | 3.9 | 3.7 | 7.6 |
| Total dur (h) | 39.8 | 36.9 | 76.7 |

**Table 2:** A breakdown of duration over train (`bbc-vad-train`) and evaluation (`bbc-vad-eval`) subsets of BBC-VAD.

## 2.5. X-vector extraction

The x-vector model [36] used in this study is based on a Kaldi recipe [37] developed for the First DIHARD Speech Diarisation Challenge [38]. It is trained on the training portion of VoxCeleb [39] and all of VoxCeleb2 [40], with Musan [41] and Room Impulse Response and Noise Database [42] used for augmentation. We adapted this baseline diarisation recipe to implement x-vector extraction. Thirty-dimensional MFCCs are extracted and passed through Cepstral Mean and Variance Normalization (CMVN), before being used as input to the x-vector model described above, with a sliding window size of 1.5 seconds and a stride of 750 milliseconds (ms), thereby producing x-vectors every 750 ms of audio signals.

## 3. PROPOSED X-VECTOR-VAD

### 3.1. X-vectors as VAD features

We first set out to visualise x-vectors extracted from `bbc-vad-eval`. We utilised scikit-learn [43] (v0.21.2) to perform dimensionality reduction. First, principal component analysis (PCA) [44] was performed on x-vectors such that the percentage variance explained post-decomposition is greater than 0.95. Secondly, t-distributed Stochastic Neighbour Embedding (t-SNE) [45] was performed on the output of PCA to visualise embeddings in a 2-dimensional space.

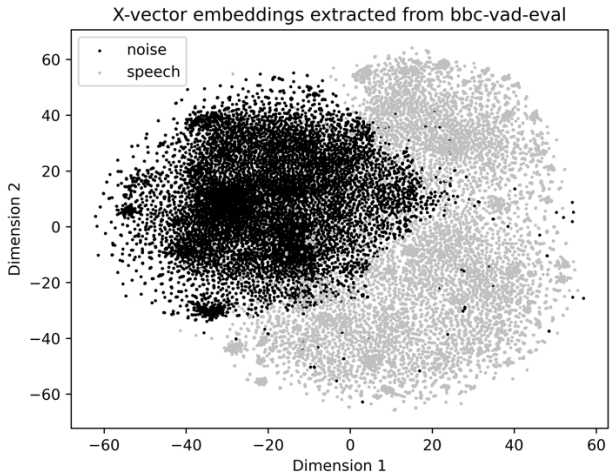

**Figure 1:** T-SNE visualisation of x-vector embeddings extracted from `bbc-vad-eval`, showing 35,702 data points labelled with either noise (black) or speech (grey).

|           |       | TPR   |       |       |       |
| Model     | FPR   | Clean | Noise | Music | All   |
|-----------|-------|-------|-------|-------|-------|
| RTC_vad   | 0.315 | 0.786 | 0.706 | 0.733 | 0.722 |
| tiny_320  | 0.315 | 0.965 | 0.826 | 0.623 | 0.810 |
| resnet_960| 0.315 | 0.992 | 0.944 | 0.787 | 0.917 |
| CNN-BiLSTM| 0.315 | 0.992 | **0.961** | **0.950** | **0.968** |
| x-vector-vad | 0.315 | **0.999** | 0.957 | 0.949 | 0.966 |

**Table 3:** TPR (true positive rate) reported at a FPR (false positive rate) of 0.315 for various VAD systems, tested on AVA-Speech. Best scores are in **bold**.

A clear separation between speech and noise clusters is observed in Figure 1, indicating that x-vectors indeed contain enough latent information to differentiate speech from noise.

### 3.2. Model description

Next, we trained a binary classifier which assigns either a speech or noise label to an input x-vector and evaluated on the AVA-Speech dataset. The classifier is a scikit-learn `Pipeline` object with a `Normalizer` followed by an estimator `LinearSVC`. The normaliser rescales each sample independently of other samples such that its L1 norm equals one. We observed empirically that this normalization of input vectors stabilised the performance of the classifier across various datasets. On top of this pipeline, probability calibration is applied using `CalibratedClassifierCV` to obtain not only the class label, but also associated probability.

### 3.3. Model performance

The `bbc-vad-train` dataset was used to fit the classifier (`x-vector-vad`) with 3-fold cross-validation. Table 3 shows the results of evaluating the `x-vector-vad` model for clean speech (Clean), speech with noise (Noise) and speech with music (Music) labels, as well as for all speech labels combined (All) from the AVA-Speech dataset. Frame-based true positive rates (TPR) for a fixed false positive rate (FPR) of 0.315 are reported. The results are shown alongside scores reported in previous studies (`RTC_vad`, `tiny_320` and `resnet_960` from [33], `CNN-BiLSTM` from [20]). `RTC_vad` generates VAD decisions over frames ranging from 10 to 30 ms, `tiny_320` and `CNN-BiLSTM` over 320 ms, `x-vector-vad` over 750 ms and `resnet_960` over 960 ms.

Our proposed `x-vector-vad` model achieves the best reported TPR of 0.999 in detecting clean speech, whilst retaining scores for detecting speech with noise and music comparable with previously reported best scores on the AVA-Speech dataset [20].

## 4. X-VECTOR-VAD FOR BROADCAST STT

Finally, we incorporated the `x-vector-vad` model into a STT system and evaluated its performance on the MGB dataset, as well as on an internal radio dataset. The STT system utilised here is based on the Kaldi `chain` setup [46]. It is a hybrid system consisting of a DNN acoustic model and a finite-state transducer (FST) language model [47]. The overview of the STT system is depicted in Figure 2.

### 4.1. Baseline STT system with WebRTC VAD

Our baseline system incorporates the publicly available implementation of VAD based on WebRTC [48][49] which outputs 30 ms frame-level decisions. It is configured to use the least aggressive filtering mode to return the maximum amount of speech. A 5-frame median filter is applied to the frame-level decisions and resulting speech segments are merged with each other if their endpoints fall within 500 ms.

As shown in Figure 3a, the WebRTC VAD is applied to the audio signal prior to x-vector extraction and clustering, which form the diarisation stage of the system and results in labelled speaker segments. These resulting segments are passed into the STT system to generate utterance-level transcripts.

### 4.2. Proposed STT system with x-vector-vad

We experiment with two approaches for the `x-vector-vad` implementation, both integrating the VAD and diarisation stages more closely than the baseline system. In both approaches, x-vectors are extracted for the entire audio stream prior to making any VAD decisions.

In the first approach (`x-vector-vad-filt`, Figure 3b), we use the trained SVM classifier to remove any noise x-vectors. The remaining speech x-vectors are then clustered into labelled speaker segments and transcribed as before. The second approach (`x-vector-vad-seg-filt`, Figure 3c) clusters all x-vectors into labelled segments first, we then use the SVM classifier to reject segments consisting of a high proportion of noise x-vectors. For both approaches, thresholds for the SVM classifier and noise x-vector proportions are discovered using the training set and chosen to maximise true positives, whilst keeping false positives at an acceptable level.

### 4.3. STT evaluation

The baseline system and the two approaches to using `x-vectors-vad` were evaluated against the MGB `eval.std` dataset which consists of 16 BBC TV programmes from a range of genres. We further tested the systems on an internal dataset consisting of 20 BBC Radio broadcasts from a range of channels and genres (`bbc-radio-eval`). This dataset adds additional diversity to our evaluation by including extra music tracks with and without vocals, speech over music and more natural conversational speech in the form of phone-ins and interviews. Neither of the test datasets annotate or exclude noise segments in the audio. This penalises systems that insert spurious transcripts for misclassified noise segments such as music, which is an undesirable behaviour for many use cases within the broadcasting industry.

Results of these evaluations are shown in Table 4. For MGB `eval.std`, x-vector level filtering (`x-vector-vad-filt`) reported the best percentage word error rage (WER) of 26.4%, whereas for `bbc-radio-eval`, segment-level filtering (`x-vector-vad-seg-filt`) resulted in the best WER score of 24.5%.

## 5. DISCUSSION

### 5.1. Latent information in x-vectors

Despite the fact that the x-vector system utilised in this study was trained specifically for the task of differentiating speakers [21][37], our analysis of x-vectors extracted from `bbc-vad-eval` shows that x-vectors retain information that differentiates speech from noise encoded in their latent space representations (Figure 1).

The speech cluster in Figure 1 is further separated into sub-clusters, some of which are tightly clustered and others more loosely. We can reasonably assume that such speech sub-clusters include speaker identities. In addition, reports from previous studies [29][30][31] indicate that x-vectors can be used to recognise various acoustic qualities in speech other than those that are useful in differentiating speakers. Further investigation into detecting metadata such as emotions and recording environments (e.g., outdoor vs. recording studios vs. over telephone) to enrich labelling of broadcast audio would be an interesting avenue to pursue.

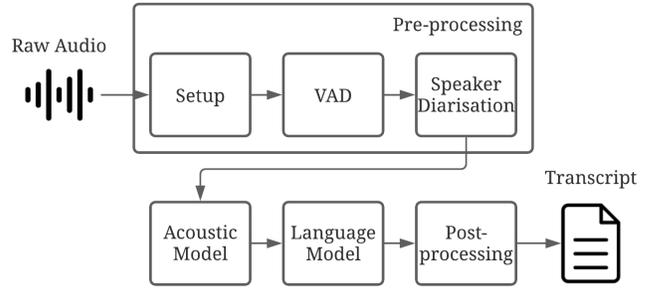

**Figure 2:** A system diagram of an end-to-end STT pipeline application based on the Kaldi `chain` setup.

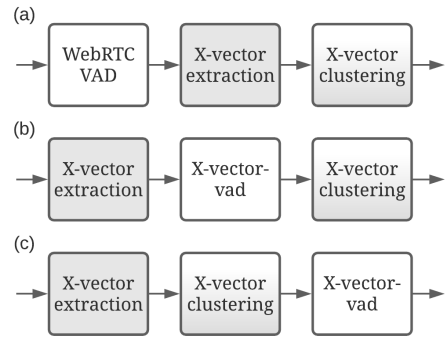

**Figure 3:** A system diagram showing a breakdown of the VAD and speaker diarisation steps. The baseline system (a) applies WebRTC VAD prior to extracting and clustering x-vectors. The proposed system with x-vector-level filtering (b) applies `x-vector-vad` prior to clustering (`x-vector-vad-filt`), whereas with segment-level filtering (c) x-vector-vad is applied after clustering (`x-vector-vad-seg-filt`).

Shifting our focus to the noise cluster in Figure 1, sub-clusters are also observed. Together with the study which showed the effectiveness of using x-vectors to classify scenes [28], we can speculate that these sub-clusters partly represent distinguishable everyday noise.

In our preliminary experiment to explore the generalisability of x-vectors as features to detect a wider range of acoustic characteristics, we also visualised x-vectors extracted from a dataset made of recordings from a BBC radio programme Desert Island Discs, in which clean speech and music borders are labelled. After applying the same dimensionality reduction procedures, the t-SNE plot showed a clear separation between speech and music clusters (data not shown). Intriguingly, the music cluster included several curved lines which are indicative of related datapoints. Could such topological features be representing music tracks? Can we detect music genres or instruments using x-vectors? Answering these questions requires further analysis.

Finally, audience reactions such as laughter and applause constitute important part of enriching the metadata of multi-genre broadcast programmes. The implied suitability of x-vectors to serve as features for a broader set of audio classification tasks warrants further investigation in this area.

| Dataset | Model | #TOT | #ERR | #INS | #DEL | #SUB | %WER |
|---|---|---|---|---|---|---|---|
| **MGB eval.std** | **WebRTC VAD** | 82361 | 22128 | 3776 | 7795 | 10557 | 26.9 |
|  | **x-vector-vad-filt** | 82364 | 21712 | 3115 | 8259 | 10338 | **26.4** |
|  | **x-vector-vad-seg-filt** | 82368 | 22012 | 2862 | 9125 | 10025 | 26.7 |
| **bbc-radio-eval** | **WebRTC VAD** | 35936 | 11144 | 5458 | 1886 | 3800 | 31.0 |
|  | **x-vector-vad-filt** | 35937 | 10206 | 4438 | 1939 | 3829 | 28.4 |
|  | **x-vector-vad-seg-filt** | 35936 | 8821 | 3021 | 2004 | 3796 | **24.5** |

**Table 4:** Evaluation results of Kaldi-based STT system with baseline (WebRTC VAD) and proposed (x-vector-vad) VAD algorithms on the MGB eval.std and bbc-radio-eval datasets. #TOT: total number of words, #ERR: total number of errors (#INS + #DEL + #SUB), #INS: number of insertion errors, #DEL: number of deletion errors, # SUB: number of substitution errors, %WER: percentage word error rate (#ERR / #TOT * 100). Best %WER scores are in **bold**.

### 5.2. X-vector based VAD

In comparing performance across various VAD algorithms, it is crucial to take note of the differences in training and evaluation datasets used. The BBC-VAD dataset used to train x-vector-vad provides more accurate speech boundaries compared to Audio Set used to train tiny_320 and resnet_960 [33]. In addition, BBC-VAD includes speech data that is more reflective of the in-the-wild broadcast materials. Furthermore, since 27% of AVA-Speech was not available at the time of analysis, this difference must be taken into account when considering results reported in Table 3.

The granularity of VAD decisions and its effect on the downstream applications should also be discussed. As each x-vector represents 750 ms of audio, x-vector-vad provides VAD decisions every 750 ms. This may not have much practical impact in the downstream STT component when used in an offline setting, as latency is less of a requirement and often speaker segments are padded at their boundaries to maximise recall. However, for online or near-real time systems, VAD algorithms such as RTC_vad [33] or CNN-BiLSTM [20], which can provide more granular decisions may be more suited.

Another aspect to consider is the computational overhead to the x-vector-vad system, as generating and clustering x-vectors is many times more expensive than a more simplistic VAD method such as RTC_vad. Again, this may be more of an issue in more resource-constrained use cases than it may be for some of the large-scale archival use cases found in the broadcasting industry. On the other hand, the use of x-vector-vad allows simplification of the speech recognition systems if the x-vector based diarisation is already in place.

### 5.3. X-vector-vad for broadcast STT

Using x-vectors as features for making VAD decisions shows an improvement in STT transcription accuracy over the baseline WebRTC VAD for both evaluation datasets chosen (Table 4). The difference is more pronounced on the radio dataset which includes more music, where the high false positive rate of the WebRTC VAD leads to the system attempting to transcribe more lyrics and hence increased insertion errors. We found that filtering noise x-vectors on an individual basis produces more stable results across both datasets, whereas filtering segments based on proportion of speech x-vectors produces significantly better results on the radio dataset, but only marginally better results on the MGB dataset alongside a marked increase in deletions.

The optimal choice of filtering method may depend on the downstream use case for the STT output. Within the broadcast industry, STT is often used within archival systems, where 100,000s of hours of content may be transcribed without any manual intervention and indexed for searching at a later date. In this case, a decrease in insertions at the expense of deletions may be a more useful trade-off, as it may ensure more correct transcripts are returned (at the expense of coverage). For use cases such as subtitling, where STT is often used as a starting point for manual editing, the number of insertions may be less of an issue — especially if they are mainly originating from music lyrics or noise being incorrectly transcribed, as these can be more easily removed by the editors.

### 6. CONCLUSION

This study introduces a novel x-vector based VAD system. We show that x-vectors extracted from broadcast audio and everyday noise retain latent information that discriminates speech from noise. The binary classifier trained with x-vectors (x-vector-vad) achieves the best reported score in detecting clean speech on the AVA-Speech dataset, whilst retaining scores for detecting speech in the presence of noise and music comparable with previously reported best scores. Importantly, the x-vector-vad model improves the accuracy of the STT system over the baseline system with WebRTC VAD on the multi-genre TV dataset (MGB) and an internal radio dataset. We conclude that x-vector-vad is a robust and accurate VAD system which enhances the quality of downstream STT transcription on a diverse range of broadcast audio materials.

## 7. ACKNOWLEDGEMENTS

We thank Fearn Bishop and Polina Proutskova for assistance in the analysis of x-vector clustering of non-speech audio. We also thank Andrew McParland, Ben Clark and Polina Proutskova for constructive comments on the manuscript and Rob Cooper for administrative support.

## 8. CODE AND DATA AVAILABILITY

The `x-vector-vad` model and the source code that implements a complete speech segmentation system are made available in the project repository (https://github.com/bbc/bbc-speech-segmenter). System requirements and usage instructions are specified in `README.md` at the root of the repository. It also provides detail on how to request access to x-vectors used for training and evaluation of `x-vector-vad`.